\documentstyle[aps,prl,twocolumn]{revtex}
\begin{document}
\title{BCS Theory for Trapped Ultracold Fermions}
\author{Georg Bruun${}^1$, Yvan Castin${}^2$, Ralph Dum${}^{1,2}$, Keith Burnett${}^1$}
\address{${}^1$ Department of Physics,
Clarendon Laboratory,
University of Oxford,
Oxford OX1 3PU,
England\\
${}^2$ Laboratoire Kastler Brossel, \'Ecole normale sup\'erieure, 
24 rue Lhomond, 75 231 Paris Cedex 5, France}
\date{22 July 1998}
\maketitle
\begin{abstract}
We develop an extension of the well-known BCS-theory to systems with trapped fermions.
The theory fully includes the quantized energy levels
in the trap. 
The key ingredient is to model the attractive interaction
between two atoms 
by a pseudo-potential which  leads to a well defined scattering problem and consequently a BCS-theory
free of divergences.
We present numerical results for the BCS critical temperature and the
temperature dependence of the gap. They are used as a test of existing
semi-classical approximations.
\end{abstract}

Considerable interest  in the field of ultracold gases
has been sparked by the achievement of Bose-Einstein condensation in the bosonic 
systems $^{87}$Rb, $^{23}$Na and $^7$Li in 1995 \cite{BEC}. 
Recently, several experimental groups have extended these  experiments to the case of 
trapped fermions.
As a first step, it is attempted to achieve a degenerate Fermi gas. In a possible
next step, the celebrated Bardeen-Cooper-Schrieffer (BCS)  phase transition could be 
observed. A promising candidate for achieving this transition is the isotope $^6$Li:
By trapping $^6$Li in two hyperfine states, one can take advantage
of the strong (attractive) interactions due to 
$s-$wave scattering between atoms in different hyperfine states. 

BCS-pairing occurs in a multitude of physical systems
(e.g.\ electrons in metals, electron-hole exciton systems and neutron-proton systems)
which share the characteristic that the formation of bound states (Cooper pairs) between 
the strongly coupled constituents
is energetically favorable \cite{deGennes}.
For ultracold atoms in traps, the interactions are much better-known than in most of the above mentioned systems \cite{FOOT}. 
Therefore, the achievement of a superfluid state in these systems opens up the possibility of 
testing our theoretical models, in particular, the validity of the BCS theory itself.
Furthermore, the interaction strength and the density of the gas are experimentally 
tunable which, in principle, makes it  possible to study 
the crossover from BCS pairing to Bose-Einstein condensation of bosonic pairs \cite{Randeria}.

To describe experiments with trapped fermions, 
one has to extend present theories (i) to include
the discrete nature of the quantum energy levels of the particles
in the trap, and (ii) to take into account the interactions specific 
to the atomic case.
We consider a model of trapped fermions with
two internal states. 
At low energies, 
the $p$-wave interaction between atoms
in the same internal state is negligible compared to the $s$-wave interaction
between atoms in different internal states. The latter interaction is characterized at 
low energies and for dilute gases
($k_F r_e<1$, where $r_e$ is the effective range
of the interaction, $k_F$ the Fermi wavevector)  
by a single parameter, the scattering length $a$.
In order to achieve pair formation the interaction has to be strong; 
here we assume $|a| \gg r_e $ \cite{Landau}. In this case, an excellent
model for the atomic interactions is provided by the pseudo-potential
discussed in \cite{Huang}. This model potential
allows us to obtain an extension of the 
BCS theory to inhomogeneous systems which is free of divergences.

We use this theory to calculate various observables including the critical BCS
temperature and the density distribution of the gas. We predict that this transition
 occurs at experimentally accessible densities and temperatures, due to the large negative
scattering length for $^6$Li atoms. We also carefully compare the results of
this general theory to those  of a theory based on the Thomas-Fermi
approximation \cite{Houbiers2} (see also \cite{baranov}). 
In that way, we establish a region of validity of this 
approximation.

Our model Hamiltonian for the trapped atomic gas includes
only interactions between atoms in different internal states:
\begin{eqnarray} \label{Ham}
\hat{H}&=&\sum_{\sigma}\int d^3r\,
 \psi_{\sigma}^{\dagger}({\mathbf{r}})
{\mathcal{H}}_0\psi _{\sigma}({\mathbf{r}}) 
\nonumber \\
&+&\frac{1}{2}\sum_{\sigma} \int d^3R d^3r d^3r'\, 
\langle {\mathbf{r}}'|\hat{V}_{RM}| {\mathbf{r}}\rangle \nonumber \\
&\times&\psi_{\sigma}^{\dagger}({\mathbf{R}}+\frac{{\mathbf{r}}'}{2})
\psi_{-\sigma}^{\dagger}({\mathbf{R}}-\frac{{\mathbf{r}}'}{2})
\psi_{-\sigma}({\mathbf{R}}-\frac{{\mathbf{r}}}{2})\psi_{\sigma}
({\mathbf{R}}+\frac{{\mathbf{r}}}
{2}).
\end{eqnarray}
Here the fermion field operator $\psi_{\sigma}({\mathbf{r}})$ annihilates 
a fermion in the position eigenstate  $|{\mathbf{r}}\rangle$
with an internal state $\sigma=+,-$. 
The single particle Hamiltonian ${\mathcal{H}}_0= -\frac{\hbar^2}{2m}\nabla^2+
U_0({\mathbf{r}})-\mu$ 
includes the trapping potential $U_0({\mathbf{r}})$.
We assume there is an equal number of particles $N_\sigma$ in each state  and 
hence a single chemical potential $\mu$\cite{MAX}.
In Eq.(\ref{Ham}), we keep a general non-local
interaction $\hat{V}_{RM}$ assuming only that it does not affect the center
of mass motion ${\mathbf{R}}$ of the interacting particles.

The interaction between atoms
is often approximated in the center of mass frame by a 
contact potential, that is
$\hat{V}_{RM}= 4\pi a\hbar^2\delta({\hat{\mathbf{r}}})/m$. However,
this approximation  leads to an ultra-violet divergent theory. This reflects the fact that the 
contact interaction is an effective low-energy interaction invalid for high energies.
One way to remove this divergence  is to introduce an  energy cut-off in the
 interaction. This approach has been 
used to describe superconductivity in metals  where a natural cut-off 
in the form of the Debye frequency exists.
Another method is to express 
the coupling constant in terms of the  two-body scattering matrix 
obtained 
from the Lippman-Schwinger equation. This renormalization scheme 
has been implemented in the literature only 
in the homogeneous case \cite{Randeria}.
Here, we put forward a technique valid also in the inhomogeneous case. We use
the pseudo-potential  $\hat{V}_{RM}$ \cite{Huang} defined for an arbitrary
function $\phi({\mathbf{r}})$ by
\begin{equation}
\langle {\mathbf{r}}|\hat{V}_{RM}|\phi\rangle\equiv g\delta({\mathbf{r}})
\partial_r[r\phi({\mathbf{r}})]
\label{PP}
\end{equation}
with  $g=4\pi a\hbar^2/m$.
We first note that in contrast to the usual contact potential
the pseudo-potential leads to a well defined two-body scattering
problem; the scattering state in the center of mass frame 
for two particles with a relative momentum
${\mathbf{p}}=({\mathbf{p}}_1-{\mathbf{p}}_2)/2$ and a relative position
$\mathbf{r}$ is
\begin{equation}
\label{SCAT}
\phi({\mathbf{r}}) = e^{i{\mathbf{p}}\cdot{\mathbf{r}}/\hbar} -{a\over 1+i pa/\hbar}
{e^{ipr/\hbar}\over r}.
\end{equation}
For a potential of finite range $r_e$ the above form, diverging
as $1/r$ for $r\rightarrow 0$, is only valid for
$r\gg r_e$ and for $p r_e/\hbar\ll  1$ \cite{Landau}. The
pseudo-potential has an effective range $r_e=0$; this does not lead
to any physical problem as the $1/r$ divergence is regularized by the 
operator $\partial_r[r\cdot]$ in Eq.(\ref{PP}). 

We follow the steady-state mean-field approach of BCS theory
\cite{VALID}:
the quartic terms of $\hat{H}$ are 
replaced by quadratic terms taking into account 
all possible binary contractions in the spirit of Wick's theorem
\cite{deGennes}.
This leads to linear equations for the field operators,
\begin{eqnarray}
i\hbar {d\over dt}\psi_{\pm}(\mathbf{r}) = {\mathcal{H}}_0 \psi_{\pm}(\mathbf{r})
&+&\int d^3r'W({\mathbf{r}},{\mathbf{r}}')\psi_{\pm}(\mathbf{r}') \nonumber
\\
&\pm&\int d^3r'\Delta({\mathbf{r}},{\mathbf{r}}')\psi^{\dagger}_{\mp}(\mathbf{r}').
\label{HB1}
\end{eqnarray}
The Hartree field $W({\mathbf{r}},{\mathbf{r}}')$ is  defined as:
\begin{eqnarray}\label{W}
W({\mathbf{r}},{\mathbf{r}}')&\equiv& \int d^3y\langle {\mathbf{r}}-{\mathbf{r}}'+
\frac{{\mathbf{y}}}{2}
|\hat{V}_{RM}|{\mathbf{r}}'-{\mathbf{r}}+\frac{{\mathbf{y}}}{2}\rangle\nonumber \\
&\times&
\langle \psi_{\pm}^{\dagger}({\mathbf{r}}'-\frac{{\mathbf{y}}}{2})
\psi_{\pm}({\mathbf{r}}-\frac{{\mathbf{y}}}{2})\rangle
\end{eqnarray}  
($W$ is the same for both internal states).
The pairing field describes correlations due to Cooper pairing: 
\begin{eqnarray}\label{PF}
\Delta({\mathbf{r}},{\mathbf{r}}')&\equiv& \int d^3y\langle {\mathbf{r}}-{\mathbf{r}}'|
\hat{V}_{RM}|{\mathbf{y}}\rangle \nonumber \\ 
&\times&
\langle \psi_{-}
\left(\frac{{\mathbf{r}}'+{\mathbf{r}}}{2}-\frac{{\mathbf{y}}}{2}\right)
\psi_{+}\left(\frac{{\mathbf{r}}'+{\mathbf{r}}}{2}+\frac{{\mathbf{y}}}{2}\right)\rangle.
\end{eqnarray}
Eq.(\ref{HB1},\ref{W},\ref{PF}) form a non-linear self-consistent problem. 
We now use 
the pseudo-potential of Eq.(\ref{PP});
as it introduces a $\delta(\mathbf{r})$, we have
$\Delta({\mathbf{r}},{\mathbf{r}'})=\delta({\mathbf{r}}-{\mathbf{r}'})\Delta({\mathbf{R}})$
and $W({\mathbf{r}},{\mathbf{r}'})=\delta({\mathbf{r}}-{\mathbf{r}'})W({\mathbf{R}})$, with ${\mathbf{R}}=({\mathbf{r}}+{\mathbf{r}'})/2$,
so that Eq.(\ref{HB1}) becomes local:
\begin{eqnarray}
i\hbar {d\over dt}\psi_{\pm}({\mathbf{R}}) = 
[{\mathcal{H}}_0 +W({\mathbf{R}})]\psi_{\pm}({\mathbf{R}})
\pm\Delta({\mathbf{R}})\psi^{\dagger}_{\mp}({\mathbf{R}}).
\label{HB2}
\end{eqnarray}
The regularizing operator $\partial_r[r\cdot]$ on the right-hand side of
Eq.(\ref{PP}) plays no role for the self-consistent
Hartree field, which is given by
$W({\mathbf{R}})\equiv 
g\langle\psi_{\sigma}^{\dagger}({\mathbf{R}})\psi_{\sigma}({\mathbf{R}})\rangle$.
Its pertinence becomes clear for the pairing field:
\begin{equation} \label{Gapdef}
\Delta({\mathbf{R}})\equiv -g\lim_{r\rightarrow 0}\partial_r
[r\langle\psi_{+}({\mathbf{R}}+\frac{{\mathbf{r}}}{2})
\psi_{-}({\mathbf{R}}-\frac{{\mathbf{r}}}{2})\rangle].
\label{DELTA}
\end{equation}
Here the operator $\partial_r[r\cdot]$ is necessary as the expectation value $\langle\psi_{+}\psi_{-} \rangle$
in Eq.(\ref{DELTA}) diverges as $1/r$ for $r\rightarrow 0$. To see this, we calculate
from Eq.(\ref{HB2}) the time 
derivative of $\langle\psi_{+}\psi_{-} \rangle$ which vanishes as the system is in a steady state:
\begin{eqnarray}
0&=&\left[-{\hbar^2\over m}[\nabla^2_{\mathbf{r}}+{1\over 4}\nabla^2_{\mathbf{R}}]+
U({\mathbf{R}}-{1\over 2}{\mathbf{r}})+U({\mathbf{R}}+{1\over 2}{\mathbf{r}})\right]
\nonumber \\ &\times&
\langle\psi_{+}({\mathbf{R}}+\frac{{\mathbf{r}}}{2})
\psi_{-}({\mathbf{R}}-\frac{{\mathbf{r}}}{2})\rangle\nonumber\\
&+& \Delta({\mathbf{R}}+{1\over 2}{\mathbf{r}})\langle\psi_{-}^\dagger({\mathbf{R}}+ 
\frac{{\mathbf{r}}}{2})
\psi_{-}({\mathbf{R}}-\frac{{\mathbf{r}}}{2})\rangle\nonumber \\
&+&
\Delta({\mathbf{R}}-{1\over 2}{\mathbf{r}})
\langle\psi_{+}^\dagger({\mathbf{R}}-\frac{{\mathbf{r}}}{2})
\psi_{+}({\mathbf{R}}+\frac{{\mathbf{r}}}{2})\rangle 
-\Delta({\mathbf{R}})\delta(\mathbf{r})
\end{eqnarray}
where $U=U_0+W$. The presence of $\delta(\mathbf{r})$ imposes a $1/r$ divergence
in the pairing field ($\nabla^2(1/r)=-4 \pi \delta(r)$): 
\begin{equation} \label{divergence}
\langle\psi_{+}({\mathbf{R}}+\frac{{\mathbf{r}}}{2})
\psi_{-}({\mathbf{R}}-\frac{{\mathbf{r}}}{2})\rangle=\frac{m}{4\pi\hbar^2r}
\Delta({\mathbf{R}})+F_{reg}({\mathbf{R}})+O(r).
\end{equation}
This $1/r$ behavior in the pairing field could actually 
be expected from the $1/r$ behavior
of the two-body scattering wavefunction Eq.(\ref{SCAT}).
Since the pseudo-potential removes this divergence, Eq.(\ref{Gapdef}) yields
$\Delta({\mathbf{R}})= -g F_{reg}({\mathbf{R}})$. 

We now compare the prediction of the present theory with the standard
theory for a homogeneous system \cite{deGennes}.
In this case $U_0\equiv 0$, and $W,\Delta$ do not depend on the position
 $\mathbf{R}$. The pairing field for a temperature $T$ can then be expressed as an integral
\begin{equation}
\langle\psi_{+}({\mathbf{R}}+\frac{{\mathbf{r}}}{2})
\psi_{-}({\mathbf{R}}-\frac{{\mathbf{r}}}{2})\rangle=
{\Delta\over 2}\int \frac{d^3k}{(2\pi)^3} {e^{i\mathbf{k}\cdot\mathbf{r}}\over E_{\mathbf{k}}}
[1-2f(E_{\mathbf{k}})]
\end{equation}
with $f(E)=[\exp(E/k_BT)+1]^{-1}$ and  $E_{\mathbf{k}}=[\Delta^2 + 
({\hbar^2k^2\over 2m}-\tilde{\mu})^2]^{1/2}$ where $\tilde{\mu}=\mu+W$.
This integral diverges for $r=0$ as in Eq.(\ref{divergence}). 
To calculate $F_{reg}$ we add and subtract from 
$\langle\psi_{+}\psi_{-}\rangle$ the integral of
a function having the same large $k$ behavior as the integrand, that
is ${1\over 2}\Delta
e^{i\mathbf{k}\cdot\mathbf{r}}/[{\hbar^2k^2\over 2m}-\tilde{\mu} 
+ i \epsilon]$,
$\epsilon\rightarrow 0+$. The integral of this function is $ G_{\mu}
(\mathbf{r})\Delta/2$ where $G_{\mu}(\mathbf{r})$ is the
single free particle Green's function.
The contribution of $G_{\mu}\Delta/2$ to $F_{reg}$ is easy to calculate
as $G_{\mu}$ is known explicitly.
The remaining integral $\langle\psi_{+}\psi_{-}\rangle-G_{\mu}\Delta/2$
converges for $r\rightarrow 0$. Using $1/(X+i\epsilon)= {\cal P}(1/X)
-i\pi\delta(X)$, we finally obtain:
\begin{equation}
F_{reg}({\mathbf{R}}) = {\Delta\over 2}
\int \frac{d^3k}{(2\pi)^3}\left[\frac{1-2f(E_{\mathbf{k}})}{E_{\mathbf{k}}}
-{\cal P}\left({1\over {\hbar^2k^2\over 2m}-\tilde{\mu}}\right)\right].
\end{equation}
The equation $\Delta = -g F_{reg}$ then coincides with the gap equation
obtained by renormalizing via the Lippman-Schwinger equation \cite{Randeria}.
 
We now turn to the inhomogeneous case, for which we
have to use a numerical approach. Here, we assume an isotropic 
harmonic
trap $U_0({\mathbf{r}})={1\over 2} m\omega^2 r^2$.
Following the Bogoliubov technique \cite{deGennes}, 
we expand the field operator in eigenmodes defined by the mode functions
$(u_\eta,v_\eta)$ solving the eigenvalue problem:
\begin{eqnarray} 
E_\eta u_\eta({\mathbf{R}}) &=& [{\cal H}_0 + W({\mathbf{R}})]  u_\eta({\mathbf{R}}) + 
\Delta({\mathbf{R}}) v_\eta({\mathbf{R}})\nonumber \\
E_\eta v_\eta({\mathbf{R}}) &=& -[{\cal H}_0 + W({\mathbf{R}})]  v_\eta({\mathbf{R}}) + 
\Delta({\mathbf{R}}) u_\eta({\mathbf{R}})\label{SYS}.
\end{eqnarray}
Here, $E_\eta$ is an elementary excitation energy and 
 $\eta$ stands for $(n,l,m)$, i.e.\  the principal quantum number $n$, the angular
momentum $l$ and the azimuthal quantum number $m$.
For thermal equilibrium, the pairing function is
\begin{eqnarray} \label{SOM}
\langle\psi_{+}({\mathbf{R}}+\frac{{\mathbf{r}}}{2})
\psi_{-}({\mathbf{R}}-\frac{{\mathbf{r}}}{2})\rangle
=\nonumber\\ \sum_{\eta}
u_{\eta}({\mathbf{R}}+\frac{{\mathbf{r}}}{2}) 
v_{\eta}({\mathbf{R}}-\frac{{\mathbf{r}}}{2})[1-2f(E_\eta)].
\end{eqnarray}
and a similar expression holds for $W$.
Inspired by the homogeneous case, we extract from
Eq.(\ref{SOM}) the term $\Delta({\mathbf{R}})/2$ times the single particle Green's 
function
\begin{eqnarray}
G_{\mu}({\mathbf{R}},{\mathbf{r}})&\equiv&\langle{\mathbf{R}}+\frac{{\mathbf{r}}}{2}|
{1\over {\cal{H }}_0}|
{\mathbf{R}}-\frac{{\mathbf{r}}}{2}\rangle\nonumber\\
&=&\frac{m}{2\pi\hbar^2r}+
G_{\mu}^{reg}(R) + O(r).
\end{eqnarray}
The key point is that the $1/r$ diverging term 
cancels in the  difference between the pairing field
and $\Delta({\mathbf{R}})G_{\mu}({\mathbf{R}},{\mathbf{r}})/2$ so that we
can take the limit  $r\rightarrow 0$. Expressing the Green's
function in the harmonic oscillator basis $\{|\phi_\eta^0\rangle\}$, we obtain
\begin{eqnarray}\label{finalGap}
\Delta({\mathbf{R}})&=& -g\sum_{\eta}\left\{
u_{\eta}({\mathbf{R}}) v_{\eta}({\mathbf{R}})[1-2f(E_{\eta})] 
\phantom{\frac{z^2}{x}}\right.\nonumber \\
&-&\left. \frac{\Delta({\mathbf{R}})}{2}
\frac{|\phi_{\eta}^0({\mathbf{R}})|^2}{(n+{3\over 2})\hbar\omega-\mu}\right\}-
\frac{g}{2}\Delta({\mathbf{R}})G_{\mu}^{reg}(R)
\end{eqnarray}
where we recall that $\eta=(n,l,m)$. In a practical numerical calculation, 
the infinite sum in Eq.(\ref{finalGap}) is, of course, replaced by a finite one. 
The regular part of the Green's function 
can be expressed in terms of a one-dimensional integral which can be
calculated numerically.

Eq.(\ref{SYS}) for the $(u_\eta,v_\eta)$'s, the self-consistent determination of 
 $\Delta({\mathbf{R}})$ given by Eq.\ (\ref{finalGap}) and the one for 
$W({\mathbf{R}})$ constitute a non-linear problem.
To obtain the critical temperature $T_c$, 
one linearizes this problem for small
$\Delta(\mathbf{R})$ which leads to the integral equation 
 $\Delta({\mathbf{s}})=\int d^3r\,M({\mathbf{s}},{\mathbf{r}})\Delta({\mathbf{r}})$. The 
temperature for which the highest eigenvalue of the kernel $M$
crosses 1, is then $T_c$ \cite{deGennes}.

We have performed a numerical diagonalization of the kernel $M$ 
for a varying chemical potential $\mu$. 
For the interaction, we took the parameters of $^6$Li that is  $a=-1140$\AA 
 \cite{Abraham} and a trapping
frequency of $820$ Hz which gives $g=-l^3 \hbar \omega$ with $l=(\hbar/m\omega)^{1/2}$.
In Fig.\ref{Tccurve}, we plot $T_c$ as a function of 
$N_\sigma$. As $\hbar\omega/k_B \simeq 40$ nK
the calculated critical temperature seems experimentally obtainable.

To obtain the spatial structure of the pairing field $\Delta(\mathbf{R})$
for arbitrary temperatures,
we solve numerically the whole self-consistent non-linear problem.
We plot in Fig.\ref{fig:gap} the pairing field
as function of $R$ for a relatively large number of particles
and a temperature much smaller than $T_c$. 
In this low temperature regime the Cooper pairing takes place 
over the whole trapped cloud.

In both of the above figures, we also compare with the Thomas-Fermi
approximation (TFA), in which the system
is treated as being locally homogeneous \cite{Houbiers2}, neglecting 
the discrete nature of the energy spectrum.
There are two conditions for the validity of the TFA. 
First, the correlation length
between the unpaired fermions $\simeq 1/k_F$ should be much shorter than the 
spatial radius $r_{TF}=\sqrt{2\mu/m\omega^2}$ of the cloud,
requiring $\mu\gg\hbar\omega$.
Also, the size $\xi$ of the Cooper pairs must be much smaller than $r_{TF}$ making a local theory 
for the pairing reasonable. For $T=0$,  
$k_F\xi\simeq\mu/\Delta_{T=0}\simeq \mu/k_B T_c$ where
we have used $\Delta_{T=0}=1.76k_BT_c$~\cite{deGennes} valid in the TFA.
For  $T=T_c$, we have the same estimate 
$k_F\xi\simeq \mu/k_BT_c$.
For any $T$ below $T_c$ the inequality $\xi\ll r_{TF}$ then reduces to 
the requirement that the critical temperature must be much larger than 
the trap level spacing for the TFA to work; i.e.\ $k_BT_c/\hbar\omega\gg 1$.

We see from Fig.1 and Fig.2 that the agreement with the TFA is reasonably good
for $k_B T_c >\hbar\omega$.
To determine the region of validity of the TFA more clearly,
we  plot in Fig.3 the quantity $S\equiv \int d^3R\ \Delta(\mathbf{R})$
as a function of temperature. In Fig.\ref{fig:size} (a), we chose a small number of 
trapped particles; in Fig.\ref{fig:size} (b) a much larger number of particles are trapped.
The temperature where the gap, that is $S$, vanishes, determines $T_c$. 
For (a), we find $k_BT_c=0.13 \hbar\omega$, for (b) $k_BT_c=2.8\hbar \omega$.
In Fig.3 (b), the agreement with the TFA is reasonably good. In Fig.3 (a),
 the prediction of the TFA is larger by two orders of magnitude;
this is not surprising given that the condition $k_BT_c\gg \hbar\omega$
is not satisfied.

In conclusion, we have implemented a 
BCS theory for a dilute gas of weakly 
interacting fermionic atoms in a trap. 
It reproduces the well-known regularized gap equation for the homogeneous 
case, and it provides a method for achieving a finite theory for a trapped gas taking into 
account the discrete nature of the normal state trap levels. 
Based on this theory, we intend in a next step to include time dependence in our treatment
to study the response of the system to external perturbations, in a search for
observable signatures of the BCS transition.

We thank A.\ J.\ Leggett, M.-O.\ Mewes, M.\ Holzmann and R.\ Combescot
for useful discussions. This work was
supported by UK-EPSRC and the TMR network EBRFMRXCT960002.
Laboratoire Kastler Brossel is a {\it unit\'e de recherche de l'Ecole
normale sup\'erieure et de l'Universit\'e Pierre et Marie Curie, associ\'ee
au CNRS}.

\begin{figure}[htb]
\vspace{6.5cm}
\caption{\label{Tccurve} The critical temperature $k_BT_c/\hbar\omega$ as a function
 of the number of particles $N_\sigma$ in each of the internal states 
for  $g=-l^3 \hbar \omega$. The solid line is obtained from  numerical solution
 of the linearized gap equation. The dashed line depicts the result of the TFA
in the parameter space where it is valid (i.e.\ $k_BT_c\gg\hbar\omega$).}
\end{figure}
\begin{figure}[htb]
\vspace{6.5cm}
\caption{\label{fig:gap} Gap function $\Delta({\mathbf{R}})/\hbar\omega$ as function
of  $R/l$ for  $g=-l^3 \hbar \omega$. We have $\mu=31.5\hbar\omega$ (yielding
 $N_\sigma\simeq 8000$) and  $k_B T=k_B T_c/10=0.28 \hbar \omega$. 
Solid line: numerical
solution of the complete BCS theory. Dashed line: TFA.}
\end{figure}
\begin{figure}[htb]
\vspace{13cm}
\caption{\label{fig:size} Size of the gap parameter  in units of $\hbar\omega l^3$
as function of $k_BT/\hbar\omega$ for a scattering length 
 $g=-l^3 \hbar \omega$, and for
(a) $\mu=11.5\hbar\omega$ yielding $N_\sigma\simeq 300$,  and (b) 
 $\mu=31.5\hbar\omega$ yielding $N_\sigma\simeq 8000$.
Solid line: numerical solution of the complete BCS theory. Dotted line
in (b): TFA.}
\end{figure}

\begin{thebibliography}{99}
\bibitem{FOOT} In the case of the cuprates and other exotic superconductors, even the 
nature of the interaction is still subject to a lively debate.
\bibitem{VALID} As the BCS theory is a mean field theory it applies here
when the pseudo-potential can be treated
in the Born approximation. This requires $p |a|/\hbar <1$;
as $p \simeq \hbar k_F$ and $k_F\simeq (3\pi^2\rho)^{1/3}$, 
the density $\rho$ at the center of the trap should satisfy
$\rho |a|^3<1/3\pi^2$. As $k_F=(2m\mu)^{1/2}/\hbar$, $\mu$
should satisfy $\mu<\hbar^2/2ma^2$.
\end{thebibliography}
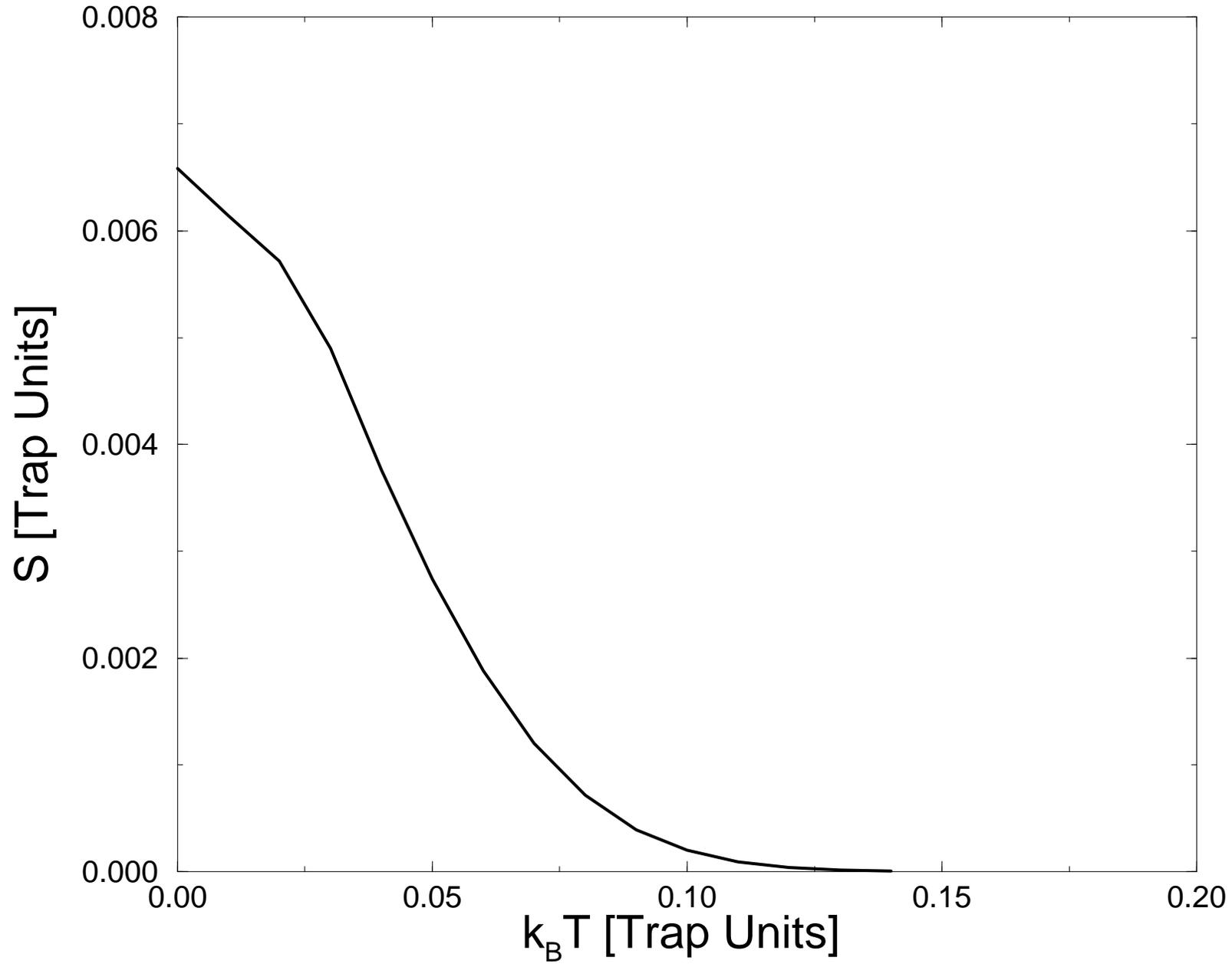
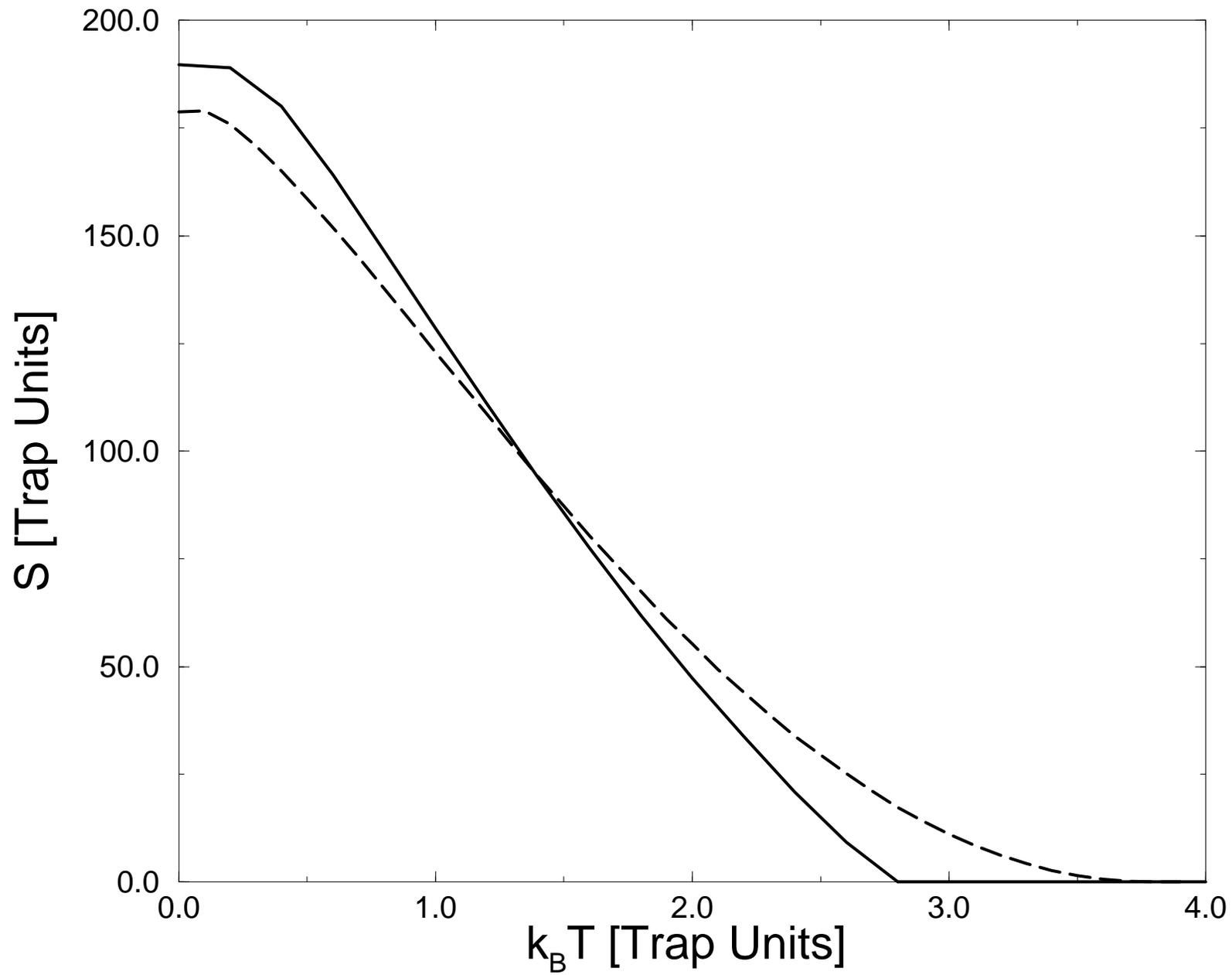
\end{document}